\documentclass[10pt,nofootinbib,twocolumn,showpacs,amsmath,amstex,amssymb,mathfonts,prl]{revtex4-2}

\usepackage{graphicx}
\usepackage{color}
\usepackage{bm}
\usepackage{amsmath}
\usepackage{amssymb}
\usepackage{amsthm}
\usepackage{amsfonts}
\usepackage[normalem]{ulem}
\usepackage[colorlinks,bookmarks=true,citecolor=blue,linkcolor=red,urlcolor=blue]{hyperref}

\usepackage{bbm}

\newcommand*{\id}{{\normalfont\hbox{1\kern-0.15em \vrule width .8pt depth-.5pt}}}

\newcommand{\be}{\begin{equation}}
\newcommand{\ee}{\end{equation}}
\newcommand{\bq}{\begin{eqnarray}}
\newcommand{\eq}{\end{eqnarray}}

\newcommand{\bs}[1]{\boldsymbol{#1}}

\usepackage{lipsum}
\theoremstyle{theorem}

\theoremstyle{theorem}

\theoremstyle{definition}

\theoremstyle{definition}

\theoremstyle{remark}

\theoremstyle{theorem}

\makeatletter
\def\@fnsymbol#1{\ensuremath{\ifcase#1\or \dagger\or \ddagger\or
   \mathsection\or \mathparagraph\or \|\or **\or \dagger\dagger
   \or \ddagger\ddagger \else\@ctrerr\fi}}
    \makeatother

\begin{document}

\title{Edge density of bulk states due to relativity}

\author{Matthew D. Horner}
\email[]{py13mh@leeds.ac.uk}
\author{Jiannis K. Pachos}
\affiliation{School of Physics and Astronomy, University of Leeds, Leeds, LS2 9JT, United Kingdom}

\date{\today}

\begin{abstract}

The boundaries of quantum materials can host a variety of exotic effects such as topologically robust edge states or anyonic quasiparticles. Here, we show that fermionic systems such as graphene that admit a low energy Dirac description can exhibit counterintuitive relativistic effects at their boundaries. As an example, we consider carbon nanotubes and demonstrate that relativistic bulk spinor states can have non zero charge density on the boundaries, in contrast to the sinusoidal distribution of non-relativistic wave functions that are necessarily zero at the boundaries. This unusual property of relativistic spinors is complementary to the linear energy dispersion relation exhibited by Dirac materials and can influence their coupling to leads, transport properties or their response to external fields.

\end{abstract}

\maketitle

{\bf \em Introduction:--} Several materials have low-energy quantum properties that are faithfully described by the relativistic Dirac equation. The celebrated example of graphene owes some of its unique properties, such as the half-integer quantum Hall effect \cite{Novoselov_2005,Novoselov_2007,Fujita} and the Klein paradox effect \cite{Katsnelson_2006,Neto}, to the relativistic linear dispersion relation describing its low-energy sector. This is by no means a singular case. A wide range of materials have been recently identified that admit 1D, 2D or 3D relativistic Dirac description, including many topological insulators and $d$-wave superconductors \cite{Wehling_2014,Moore, Jia, Hasan, Hasan2, Bernevig}. The unusual dispersion relation of Dirac materials gives rise to effective spinors, where the sublattice degree of freedom is encoded in the pseudo-spin components. Nevertheless, the emerging excitations are spinor quasiparticles that can exhibit novel transport properties or responses to external fields akin only to relativistic physics \cite{Neto}.

Here we present another counter-intuitive aspect of relativistic physics in Dirac materials manifested by the behaviour of bulk states at the boundaries. In general, the choice of boundary conditions one imposes on single-particle wavefunctions of a system must ensure its Hamiltonian remains Hermitian. For the example of a non-relativistic particle in a box obeying the Schr\"odinger equation, the boundary conditions are simply that the wavefunction vanishes on the walls of the box. However, for spin-$1/2$ particles of mass $m$ obeying the $(2+1)$D Dirac equation 
\be
\left( -i \alpha^i \partial_i  + \beta m \right) \psi(\boldsymbol{r}) = E\psi(\boldsymbol{r}),  \,\,\psi(\boldsymbol{r}) = \begin{pmatrix} \psi_{\uparrow}(\boldsymbol{r}) \\ \psi_\downarrow(\boldsymbol{r}) \end{pmatrix},
\label{eq:Dirac}
\ee
where $\alpha^i$ and $\beta$ are the $2 \times 2$ Dirac alpha and beta matrices, vanishing of the spinor $\psi(\boldsymbol{r})$ is not possible on all boundaries without the solution being trivially zero everywhere. The requirement that the Dirac Hamiltonian $h = -i \alpha^i \partial_i + \beta m$ is Hermitian with respect to the inner product $\int_D \mathrm{d}^2 r \psi^\dagger(\boldsymbol{r}) \phi(\boldsymbol{r})$ on a finite domain $D$ is that the charge current $J^i(\boldsymbol{r}) = \psi^\dagger(\boldsymbol{r}) \alpha^i \psi(\boldsymbol{r})$ normal to the boundary $\partial D$ is zero for all spinors. In other words, if $\hat{\boldsymbol{n}} $ is the outward pointing normal to the boundary, then
\begin{equation}
\hat{\boldsymbol{n}}(\boldsymbol{r}_0) \cdot \boldsymbol{J}(\boldsymbol{r}_0) = 0 \label{eq:boundary_flux}
\end{equation}
for all points $\boldsymbol{r}_0 \in \partial D$~\cite{Berry}. This condition ensures that particles are trapped in $D$. In contrast to the non-relativistic case, the zero flux condition of Eq.  (\ref{eq:boundary_flux}) allows for bulk solutions $\psi(\boldsymbol{r})$ whose charge density $\rho(\boldsymbol{r}_0) = \psi^\dagger(\boldsymbol{r}_0) \psi(\boldsymbol{r}_0)$ is non-zero on the boundaries~\cite{Alberto,Alonso}.

To exemplify our investigation, we consider how bulk spinor states behave at the edges of a zig-zag carbon nanotube -- a system which is described by the Dirac equation of Eq. (\ref{eq:Dirac}). We find that bulk states have support on the edges of the nanotube depending on the size of the system. Importantly, these relativistic effects become more dominant for gapless nanotubes, corresponding to systems with a multiple of three unit cells in circumference, or when the length of the nanotube is small. Such relativistic properties of spinor eigenstates are expected to be present in all Dirac-like materials and are complementary to the typically linear dispersion relation they exhibit. Bulk states with non-zero density at the boundaries are expected to impact the coupling of Dirac materials to external leads, their transport properties or their response to external magnetic fields. 

\begin{figure}
\centering
$\vcenter{\hbox{\includegraphics[width=0.59\columnwidth]{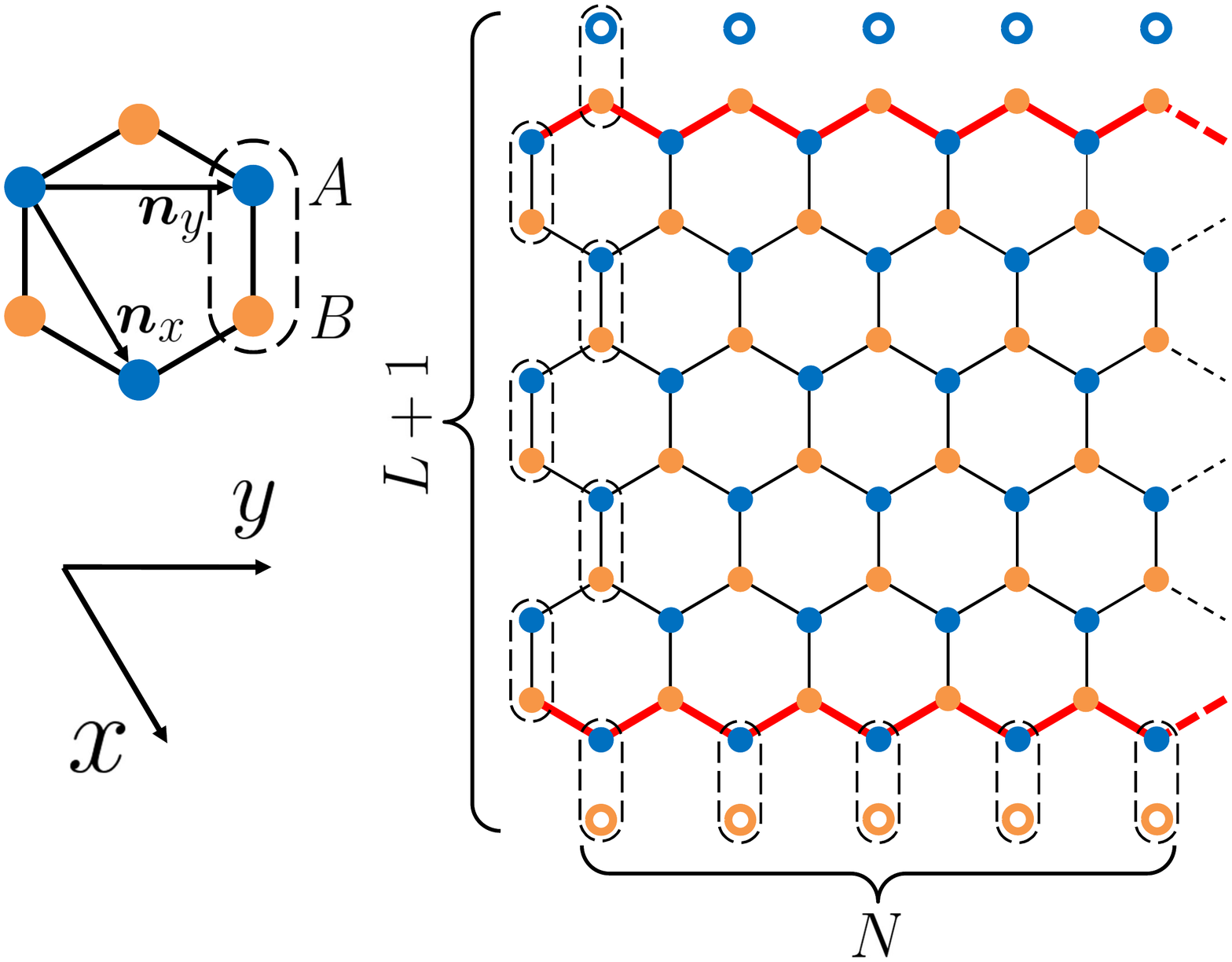}}}$
$\vcenter{\hbox{\includegraphics[width=0.39\columnwidth]{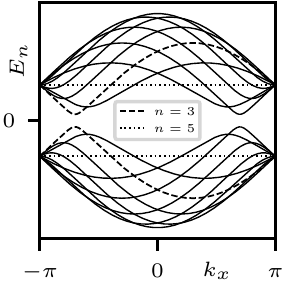}}}$
\caption{(Left) The honeycomb lattice comprising of two triangular sublattices $A$ and $B$. The lattice basis vectors $\boldsymbol{n}_x = \frac{1}{2}(1,-\sqrt{3})$ and $\boldsymbol{n}_y = (1,0)$ are depicted with corresponding non-Cartesian coordinates $x$ and $y$. A nanotube with zig-zag boundaries (red lines) is depicted with length $L$ corresponding to $L+1$ unit cells (dashed ovals) in the $\boldsymbol{n}_x$ direction while the $\boldsymbol{n}_y$ direction is periodic with circumference $N$. The boundary condition is given by the top $A$ sites and bottom $B$ sites having zero population. (Right) The band structure of a zig-zag nanotube of circumference $N = 10$ is displayed, where $n=3$ (dashed lines) is one of the two minimum gap bands. All bands have a single minimum except $n = N/2 = 5$ which is completely flat (dotted line).}
\label{fig:honeycomb}
\end{figure}

{\bf \em Relativistic description of zig-zag carbon nanotubes:--} The honeycomb lattice of graphene is formed from two triangular sublattices $A$ and $B$. We take the two basis vectors $\boldsymbol{n}_x$,$\boldsymbol{n}_y$  and we take the vertical links as our unit cells, as shown in Fig. \ref{fig:honeycomb}. We label our lattice sites with the pair $(\boldsymbol{r},\mu)$, where $\boldsymbol{r} = x \boldsymbol{n}_x + y \boldsymbol{n}_y$ labels the position of the unit cell with non-Cartesian coordinates $ x,y \in \mathbb{Z}$, while $\mu \in  \{A,B \}$ labels the site within the unit cell. The Hamiltonian of the system is given by $
H=-t\sum_{\langle \bs r , \bs r' \rangle} a^\dagger_{\bs r} b_{\bs r'} + \text{h.c.}$, where $a^\dagger_{\bs r}$ ($b^\dagger_{\bs r}$) creates a fermion on sublattice $A$ ($B$) of unit cell ${\bs r}$~\cite{Neto}. Bloch momenta are given by $\boldsymbol{k} = \frac{1}{2 \pi} ( k_x \boldsymbol{G}_x + k_y \boldsymbol{G}_y)$, where $\boldsymbol{G}_x$, $\boldsymbol{G}_y$ are the reciprocal basis vectors and  $k_x,k_y \in [- \pi, \pi]$ are the corresponding coordinates of the Brillouin zone (BZ) (see Appendix).

To study the low-energy properties of a finite zig-zag nanotube, we first take the continuum and thermodynamic limit in the the $\boldsymbol{n}_x$ direction only, whilst keeping the the periodic $\boldsymbol{n}_y$ direction finite and discrete, with $N$ unit cells in circumference. This gives rise to $N$ bands parametrised by momenta $k_y  = 2 n \pi/N$, where $n$ is an integer~\cite{Charlier,Saito}. The $n$th band has the one-dimensional dispersion relation
\begin{equation}
E_{n}(k_x) = \pm t \sqrt{3 + 2g_n(k_x)}, \label{eq:dispersion}
\end{equation}
where $g_n(k_x) =  \cos \left( \frac{2 n \pi}{N} \right) +  \cos \left( \frac{2 n \pi}{N} - k_x \right) +  \cos (k_x)$.
The zig-zag nanotube is typically gapped, unlike an infinite flat sheet of graphene which is gapless. Each conduction band contains a single minima, as seen in Fig.~\ref{fig:honeycomb}, which dictates the low-energy physics for that particular band. Our model is a simplified version of a carbon nanotube as we ignore effects due to curvature and spin orbit coupling that are not relevant to our investigation~\cite{Kane,Ando,Marganska,Marganska2,Akhmerov,Neto,Saito}.

Following the literature~\cite{Marganska,Akhmerov}, we expand the Hamiltonian about the minima of Eq. (\ref{eq:dispersion}) by letting $k_x=K_\text{min}+p$, for each band $n$, yielding the $(1+1)$D massive Dirac Hamiltonian $H_n = \int \mathrm{d}x \psi^\dagger_{n} h_n\psi_{n}$, with
\begin{equation}
h_n =  -i e_n \sigma^y \partial_x + \Delta_n \sigma^x , \,\, \psi_n(x) = \begin{pmatrix} a_n(x) \\ b_n(x) \end{pmatrix},
\label{eq:ham}
\end{equation}
where the sublattices $A$ and $B$ of the unit cell are encoded on the pseudo-spin components, $e_n  = 2t \cos \left( \frac{n \pi}{N} \right)$ is the spatial component of the zweibein and $\Delta_n$ is the energy gap of the $n$th band, given by
\begin{equation}
\Delta_n  =  t\left( 2 \cos \left( \frac{n \pi}{N} \right) - 1 \right). \label{eq:gap} 
\end{equation}
See Appendix A for a derivation. We now truncate the length of the nanotube to a finite length $L$. We construct standing waves $\psi(x) = (\psi_A(x),\psi_B(x))^\text{T}$ from forward and backward propagating eigenstates of $h_n$ of Eq. (\ref{eq:ham}). The zig-zag boundary conditions are $\psi_A(0) = \psi_B(L) = 0$, where $x=0$ and $x=L$ are the coordinates of the unit cells of the top and bottom boundaries, as shown in Fig.~\ref{fig:honeycomb}~\cite{Marganska,Marganska2}. These conditions obey the zero flux condition of Eq. (\ref{eq:boundary_flux}). This gives the solutions
\begin{subequations}
\begin{align}
\psi_{n,p}(x)  & = \mathcal{N}_{n,p} \begin{pmatrix}  \sin(p x)  \\\sin(p x + \theta_{n,p}) \end{pmatrix} , \label{eq:wavefunctions} \\
\theta_{n,p}  & = \arg(\Delta_n + ie_np),  \label{eq:phase}
\end{align}
\end{subequations} 
where $\mathcal{N}_{n,p}$ is a normalisation constant and $\theta_{n,p}$ is a relative phase shift between the $A$ and $B$ sublattice wavefunctions. The quantised momenta $p$ are solutions to the transcendental equation
\begin{equation}
pL + \theta_{n,p} = m \pi , \ m \in \mathbb{N}, \label{eq:momentum_eqn}
\end{equation}
which can be solved numerically (see Appendix A). Note that the wavefunctions of Eq.~(\ref{eq:wavefunctions}) correspond to bulk states, however graphene with zig-zag boundaries also supports zero-energy states localised at the edges~\cite{Bernevig}. Edge states correspond to complex solutions of Eq. (\ref{eq:momentum_eqn}) and are not considered here~\cite{Neto,Marganska,Marganska2}. 

{\bf \em Relativistic edge effects of bulk states:--} The $U(1)$ electric charge density of $(1+1)$D Dirac spinors $\psi = (\psi_A,\psi_B)^T$  is given by $\rho = \psi^\dagger \psi = |\psi_A|^2 + |\psi_B|^2$. With our interpretation of the pseudo-spin components $\psi_A(x)$ and $\psi_B(x)$ as the sublattice wavefunctions, where $x$ labels the unit cell, $\rho$ is therefore the charge density with respect to the unit cells. For the bulk standing wave solutions of Eq. (\ref{eq:wavefunctions}), we have
\begin{equation}
\rho_{n,p}(x) = |\mathcal{N}_{n,p}|^2 \left( \sin^2(p x) + \sin^2(p x + \theta_{n,p}) \right), 
\label{eq:rho}
\end{equation}
which gives a charge density at the boundaries of
\begin{equation}
\rho_{n,p}(0) = \rho_{n,p}(L) = |\mathcal{N}_{n,p}|^2 \sin^2(\theta_{n,p}). \label{eq:edge_density}
\end{equation}
We see that it is possible to have $\rho_{n,p}(0)\neq 0$ due to the phase difference, $\theta_{n,p}$, which is purely a relativistic effect. 

The edge charge density of bulk states is maximal when $\theta_{n,p} = \pm \pi/2$. Referring to Eq. (\ref{eq:phase}), this is achieved when $\Delta_n = 0$, i.e., when the $n$th band is gapless. From Eq. (\ref{eq:gap}) we see that the gap closes if $n/N = \pm 1/3$ which is only possible if $N$ is a multiple of three. Note that, for a gapless band, the charge density of Eq. (\ref{eq:rho}) is also completely uniform with
\begin{equation}
\rho_{n,p}(x) = \frac{1}{L},  
\label{eq:edge_density2}
\end{equation} 
which is independent of the momentum $p$, where we have chosen a $1$D normalisation. On the other hand, when the system is gapped, then the density oscillates along the length of the nanotube and becomes vanishingly small at the edges. This shift in behaviour of the charge density reflects the expected transition from the relativistic to non-relativistic regime witnessed in confined Dirac particles as their mass increases~\cite{Alberto}. 

\begin{figure}[t!]
\includegraphics[width=\columnwidth]{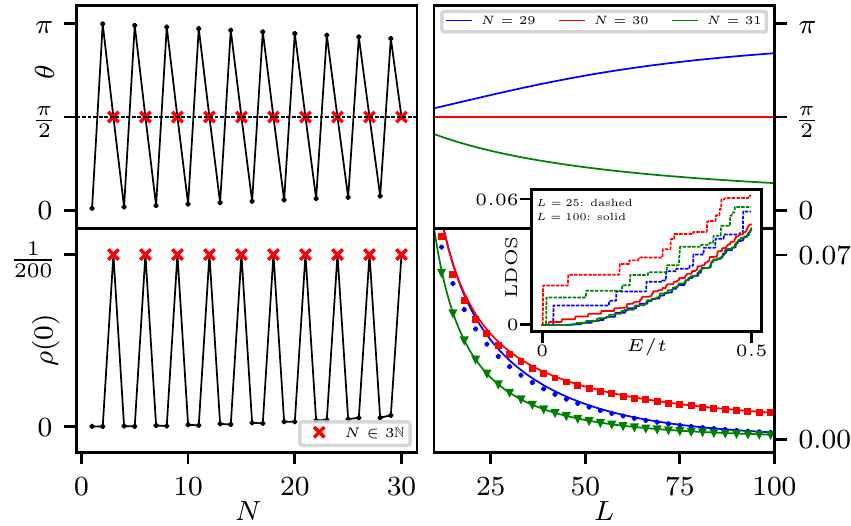}
\caption{(Left) The phase shifts $\theta \equiv \theta_{n,p}$ of Eq. (\ref{eq:phase}) and the numerically measured edge densities $\rho(0) \equiv \rho_{n,p}(0)$ versus circumference $N$ for the ground state of a system of fixed length $L = 200$. When $N$ is a multiple of three, i.e., when the system is gapless, the phase shift is $\theta =\pi/2$ and the edge density is $\rho(0)=1/L = 0.005$ confirming Eq. (\ref{eq:edge_density2}). (Right) The phase shifts $\theta$ and edge densities $\rho(0)$ versus system length $L$ for the ground state of the gapless systems $N=30$ and its two neighbouring gapped systems $N=29$ and $N=31$. The solid line represents the analytical formulas whilst the points represent numerics. The edge density for the gapless system $N=30$ goes as $1/L$, while for gapped systems $N=29$ and $N=31$ the edge density tends to zero quickly in agreement with Eq. (\ref{eq:edge_density}). (Inset) The integrated LDOS at the edge $x = 0$ for a nanotube of circumferences $N=29,30,31$ and lengths $L=25$ (dashed lines) and $L=100$ (solid lines). The LDOS displays the predicted behaviour of maximising for gapless systems ($N= 30$) and increasing with smaller system size $L$.}
\label{fig:phases_densities}
\end{figure}

The stark contrast between gapped and gapless systems is confirmed numerically (see Appendix B for numerical details). The left-hand column of Fig. \ref{fig:phases_densities} shows the edge density $\rho$ of the ground state of a system of length $L = 200$ for varying circumferences $N$. When $N$ is a multiple of three, i.e., when the system is gapless, the edge density spikes to the expected value of $1/L = 0.005$. On the other hand, when $N$ is not a multiple of three, i.e., when the system is gapped, the edge density is small. This behaviour is a consequence of the highly oscillating phase shift $\theta$. When $N$ is a multiple of three, the phase shift is $\pi/2$ exactly, maximising the edge density according to Eq. (\ref{eq:edge_density}). The smaller $N$ is, the stronger the effect as the difference between gapped and gapless systems is much stronger due to the gaps being larger. However, as $N$ increases, all zig-zag nanotubes tend towards gapless systems even if $N$ is not a multiple of three, as there exists a band $n$ such that $n/N \approx \pm 1/3$ when $N$ is large, so the gap of Eq. (\ref{eq:gap}) begins to close, so all systems begin to behave similarly.

The relativistic boundary effects also have a system length dependence~\cite{Alberto}. The right-hand column of Fig. \ref{fig:phases_densities} shows the numerically measured edge density $\rho$ of the ground state of a the gapless system $N=30$ and its two neighbouring gapped systems $N=29$ and $N=31$ for varying system lengths $L$. The edge density of the gapless system $N=30$ goes as $1/L$ whereas the edge density gapped systems $N=29$ and $N=31$ tends to zero quickly, both in accordance with Eq. (\ref{eq:edge_density}). It is worth noting that, despite the fact that the analytic results have been derived in the large $L$ limit where the continuum approximation holds, the numerics and analytics are in surprisingly good agreement even for very small $L$. This verifies the theoretically predicted relativistic effects of nanotubes with small length $L$ where the violation of the non-relativistic zero edge density is expected. Note that this behaviour repeats itself for any $N$ that is a multiple of three and its two neighbouring sizes above and below, which the left hand side of Fig. \ref{fig:phases_densities} demonstrates.

To explain the system size dependence of the charge density, note that for very small $L$ the allowed momenta $p$ satisfying Eq. (\ref{eq:momentum_eqn}) become very large. In this case, the imaginary contribution to the phase $\theta_{n,p}  = \arg(\Delta_n + ie_np)$ dominates, giving $\theta_{n,p}\approx \pi/2$ even if the gap is non-zero, as seen in the right-hand column of Fig. \ref{fig:phases_densities}. Hence, the edge density of Eq. (\ref{eq:edge_density}) becomes significant for small system sizes. For the gapless case, the phase is exactly equal to $\pi/2$ regardless of the value of $p$ or system size $L$. This yields a uniform charge density throughout the nanotube, resulting in the $1/L$ edge density as observed.

Finally, Fig. \ref{fig:phases_densities} shows the integrated local density of states (LDOS) on the edge at $x=0$ given by $N(E,\boldsymbol{r}) = \sum_m \rho_m(\boldsymbol{r}) \Theta(E-E_m)$, where $\rho_m(\boldsymbol{r})$ is the unit cell charge density of the $m$th eigenstate of the $2$D model with eigenvalue $E_m$. We present this for systems $N=29,30,31$ and $L= 25,100$. The edge LDOS is maximised for a fixed $L$ when the system is gapless, so for $N=30$ in this case. Moreover, the LDOS increases as the system size decreases, which provides a clear signature for the observation of the relativistic edge effect.

To summarise, the edge density is prominent if either the system is gapless, so $N$ is a multiple of three, or the system length $L$ is small. The typical lattice constant of a nanotube is given by $|\boldsymbol{n}_x| = |\boldsymbol{n}_y| \approx 2.46$\AA ~\cite{Altland,Saito}, so Fig. \ref{fig:phases_densities} applies to systems on the order of $1$nm in diameter and $10$nm in length. However, the dependence on whether the system is gapless or not is very strong, so this effect holds for much larger circumferences $N$ and lengths $L$. Therefore, we expect these results to hold for a wide range of experimentally accessible sizes.

{\bf \em Relativistic spinors from non-relativistic wavefunctions:--} To explain the emergence of relativistic boundary effects from a non-relativistic model, we focus on the sublattice wavefunctions $\psi_A$ and $\psi_B$. For concreteness, we examine a nanotube of dimension $(N,L) = (30,200)$ and $(N,L) = (31,200)$ which have gapless and gapped spectra, respectively. 

\begin{figure}[t!]
\includegraphics[width=\columnwidth]{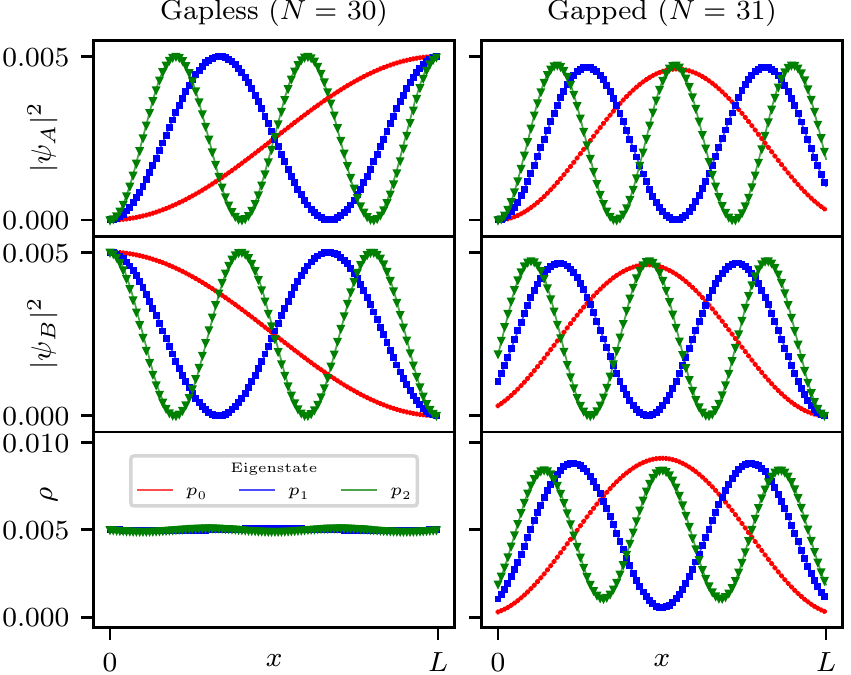}
\caption{(Left) A comparison of the analytical wavefunctions $|\psi_A|^2$ and $|\psi_B|^2$ of Eq. (\ref{eq:wavefunctions}) and charge densities $\rho$ of Eq. (\ref{eq:edge_density}) to the numerical simulation (markers) for the gapless system $(N,L) = (30,200)$. We present the first three excited states above the Fermi energy $E=0$. We observe the wavefunctions act highly-relativistically, with a large boundary support and uniform charge density of $1/L = 0.005$. (Right) We present the same information for the gapped system $(N,L) = (31,200)$. The behaviour contrasts highly with the gapless system despite $N$ only being greater by one. The wavefunctions and densities display a Schr\"odinger-like profile with a much smaller edge density that tends to zero as $L$ increases as seen in Fig. \ref{fig:phases_densities}.}
\label{fig:wavefunctions}
\end{figure}

In the left-hand column of Fig.~\ref{fig:wavefunctions} we compare the numerical sublattice wavefunctions $\psi_A(x)$, $\psi_B(x)$ and the charge densities $\rho(x)$ to the analytical results of Eq. (\ref{eq:wavefunctions}) and Eq. (\ref{eq:rho}) respectively, for the first three excited states above the Fermi energy for the gapless system $(N,L) = (30,200)$. We see that the sublattice wavefunctions $\psi_A$ and $\psi_B$ are highly out of phase and maximise the edge support at $x=L$ and $x=0$ respectively, yielding a charge density $\rho(x)$ with minor oscillations about the predicted uniform value of $ 1/L = 0.005$. These oscillations are caused by finite-size effects. 

In the right-hand column of Fig \ref{fig:wavefunctions}, we present the same information for the gapped system $(N,L) = (31,200)$. Despite $N$ increasing only by $1$, the fact the system now has a gap results in wavefunctions $\psi_A(x)$ and $\psi_B(x)$ that contrast considerably to the gapless case, with a charge density $\rho(x)$ that displays a more Schr\"odinger-like oscillatory profile. As the system size $L$ increases, the relative phase shift $\theta$ modolo $\pi$ between $\psi_A(x)$ and $\psi_B(x)$ decreases, as seen in Fig. \ref{fig:phases_densities}, and the wavefunctions begin to display the Schr\"odinger-like profile that tends to zero on the boundaries. However, this is not the case for  gapless systems as the phase shift is always $\pi/2$ regardless of system size, as seen in Fig. \ref{fig:phases_densities}.

\begin{figure}[t!]
\includegraphics[width=\columnwidth]{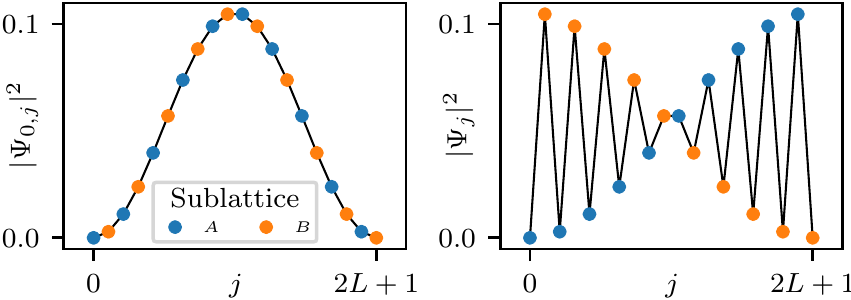}
\caption{The full single-particle ground state $\Psi_{0,j}$ and the first state above the Fermi energy $\Psi_j$ for a system of size $(N,L) = (30,9)$. The emergent relativistic physics near the Fermi energy can be seen clearly as a result of aliasing of a high frequency Schr\"odinger wavefunction. Comparing this with the left-hand column of Fig. \ref{fig:wavefunctions}, we see how the sublattice wavefunctions $\psi_A$ and $\psi_B$ described by the spinor Eq. (\ref{eq:wavefunctions}) emerges.}
\label{fig:full_wavefunctions}
\end{figure}

We now analyse the total wavefunctions $\Psi_j$ of the lattice fermions, where $j \in \mathbb{N}$ is the real space coordinate of the bipartite lattice, alternating between sublattices $A$ and $B$. This coordinate should be contrasted to the unit cell coordinate $x$ of the spinor $\psi(x)$. Fig. \ref{fig:full_wavefunctions} shows the wavefunctions of the single-particle eigenstate with the most negative energy below the Fermi energy, $\Psi_{0,j}$, and the first single-particle eigenstate above the Fermi energy, $\Psi_j$, for a system of dimension $(N,L) = (30,9)$. 

The wavefunctions $\Psi_{0,j}$ and $\Psi_j$ are both non-relativistic wavefunctions which vanish on the boundaries. This is to be expected as the microscopic model is non-relativistic. However, due to high frequency oscillations, the support of $\Psi_j$ on each sublattice is highly out of phase. Comparing with the left-hand column of Fig. \ref{fig:wavefunctions}, we see that these oscillations give the impression of two separate wavefunctions faithfully described by the components of a Dirac spinor. Non-relativistically, we expect the system to behave like a particle in a box, so we take the ansatz wavefunction  $\Psi_j \propto \sin(pj)$. From inspection, we see that this matches the numerics for momenta $p = (l+1)\pi/2l$, where $l = 2L+1$ is the total length of the bipartite chain and $L+1$ is the number of unit cells as defined in Fig. \ref{fig:honeycomb}. This gives a wavelength comparable to the lattice spacing. Therefore, the emergent relativistic physics described by the spinor of Eq. (\ref{eq:wavefunctions}) is a consequence of aliasing from sampling a high frequency non-relativistic wavefunction at discrete intervals. This effect is independent of length $L$. Such high frequency wavefunctions correspond to the middle of the spectrum where the relativistic linear dispersion is present. 

On the other hand, gapped systems display a Schr\"odinger-like wavefunction for both $\Psi_{0,j}$ and $\Psi_j$ if the system length $L$ is large. This can be seen clearly in the right-hand column of Fig. \ref{fig:wavefunctions} where the sublattice wavefunctions are almost in phase. The total wavefunctions that describe these can also be described by the ansatz wavefunction of a particle in a box, but for a small $p$ instead, so $A$ and $B$ sublattices are now more in phase, similar to the left hand side of Fig. \ref{fig:full_wavefunctions}. We also see this in the right hand side of Fig. 2 where the edge densities drop to zero on the walls, suggesting a non-relativistic behaviour.

{\bf \em Conclusion:--} Our analysis demonstrates that relativistic effects can dominate certain geometries of Dirac materials, resulting in large edge support. We studied this effect analytically and numerically for zig-zag carbon nanotubes and demonstrated that it holds strongly for a wide range of experimentally accessible sizes. We found that the effect is dominant when the system is either gapless or has a small length on the order of $10$nm. Nevertheless, this relativistic effect is general and it is expected to be present in 1D, 2D and 3D materials with the same qualitative properties presented here. While high edge densities of bulk states should be measurable with STM \cite{Andrei, Kim,Venema,Hassanien}, it is expected to have a significant effect on the conductivity of the material when attaching leads to its boundaries or its response to a magnetic field \cite{Laird,Marganska,Marganska2,Hiroshi}. In addition, determining if such effects will be present in 2D materials containing a finite density of defects which effectively imposes boundary conditions on the wavefunctions within the material will be intriguing \cite{Algharagholy,Araujo,Neto,Dutreix}. We leave these questions for a future work. 

\begin{acknowledgments}
{\bf \em Acknowledgements:--} We would like to thank Oscar Cespedes, Jamie Lake, Alex Little and Satoshi Sasaki for inspiring conversations. This work was supported by the EPSRC grant EP/R020612/1. Statement of compliance with EPSRC policy framework on research data: This publication is theoretical work that does not require supporting research data.
\end{acknowledgments}

\bibliography{graphene_refs}
\clearpage
\onecolumngrid

\section{Appendix A: Continuum limit of zig-zag carbon nanotubes}

The honeycomb lattice of graphene is formed from a triangular Bravais lattice with a unit cell containing two sites, one on sublattice $A$ and the other on sublattice $B$, as shown in Fig. 1 of the main text. The Bravais lattice is generated by the two basis vectors
\begin{equation}
\boldsymbol{n}_x = \frac{a}{2}(1,-\sqrt{3}), \ \boldsymbol{n}_y = a(1,0),
\end{equation}
where $|\boldsymbol{n}_x| = |\boldsymbol{n}_y| = a$ is the lattice spacing. In this Letter, we take $a=1$, but we leave it in this supplementary material for completeness. We label our lattice sites with the pair $(\boldsymbol{r},\mu)$, where $\boldsymbol{r} = x \boldsymbol{n}_x + y \boldsymbol{n}_y$ labels the position of the unit cell which we take to coincide with sublattice $A$, where $x,y \in \mathbb{Z}$ are the \textit{non-Cartesian} coordinates, and $\mu \in \{ A,B \}$ labels the site within the unit cell. The corresponding reciprocal basis is given by
\begin{equation}
\boldsymbol{G}_x = \frac{2 \pi}{a\sqrt{3}}(0,-2), \  \boldsymbol{G}_y = \frac{2 \pi}{a\sqrt{3}}(\sqrt{3},1).
\end{equation}
where $\boldsymbol{n}_i \cdot \boldsymbol{G}_j = 2\pi \delta_{ij}$. With this reciprocal basis, the Bloch momenta are given by $\boldsymbol{k} = \frac{a}{2 \pi}(k_x \boldsymbol{G}_x + k_y \boldsymbol{G}_y)$, where $k_x , k_y \in [- \pi/a, \pi/a]$ defines the Brillouin zone (BZ) which is square in this coordinate system. The components of momenta in the $\boldsymbol{n}_i$ direction are given by $k_i = \boldsymbol{k} \cdot \hat{\boldsymbol{n}}_i$, where $\hat{\boldsymbol{n}}_i = \boldsymbol{n}_i/a$ are unit vectors in these directions.

\begin{figure}[h]
\centering
\includegraphics[width=0.5\columnwidth]{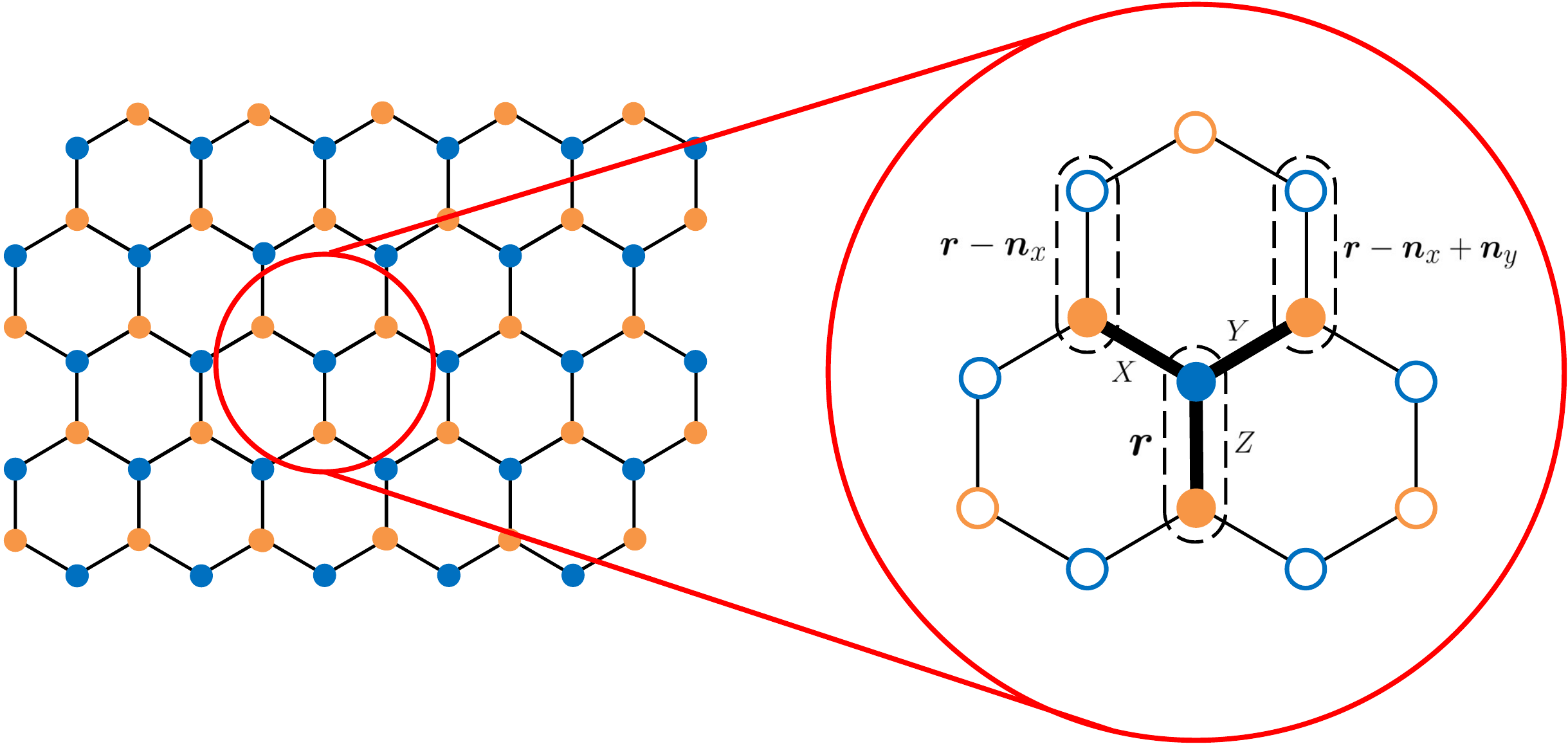}
\caption{The nearest-neighbour hoppings of the Hamiltonian can be tiled across the honeycomb with this particular unit cell of hoppings along $X$, $Y$ and $Z$ links represented by the thick black links. The unfilled circles represent sites that are not contained in this choice of unit cell.}
\label{fig:unit_cell}
\end{figure}

We take the tight-binding Hamiltonian of graphene with nearest-neighbour hoppings only. With our choice of unit cell, basis vectors and labelling convention, the tight-binding Hamiltonian can be written as
\begin{equation}
\begin{aligned}
H & = - t \sum_{\langle \boldsymbol{r} , \boldsymbol{r}' \rangle } a^\dagger_{\boldsymbol{r}} b_{\boldsymbol{r}'} + \text{h.c.} \\
& = -t \sum_{{\bs r} \in A} a^\dagger_{\bs r} \left( b_{\bs r} + b_{\bs r - {\bs n}_x} + b_{\bs r - {\bs n}_x + {\bs n}_y} \right) + \text{h.c.} \label{eq:ham_tb}
\end{aligned}
\end{equation}
where $t$ is the hopping parameter and $a^\dagger_{\boldsymbol{r}}$ ($b^\dagger_{\boldsymbol{r}}$) are fermionic operators which create an electron on sublattice $A$ ($B$) of the unit cell located at $\boldsymbol{r}$ \cite{Neto}. The second equality corresponds to the fact one can construct the honeycomb lattice by tiling sublattice $A$ with the ``Y" shape of links that originate from a single $A$ site. Careful consideration must be taken with the top and bottom row when we have boundaries as the external vertical links are missing, as shown in Fig. \ref{fig:unit_cell}.

\begin{figure}
\centering
\includegraphics[width=0.5\textwidth]{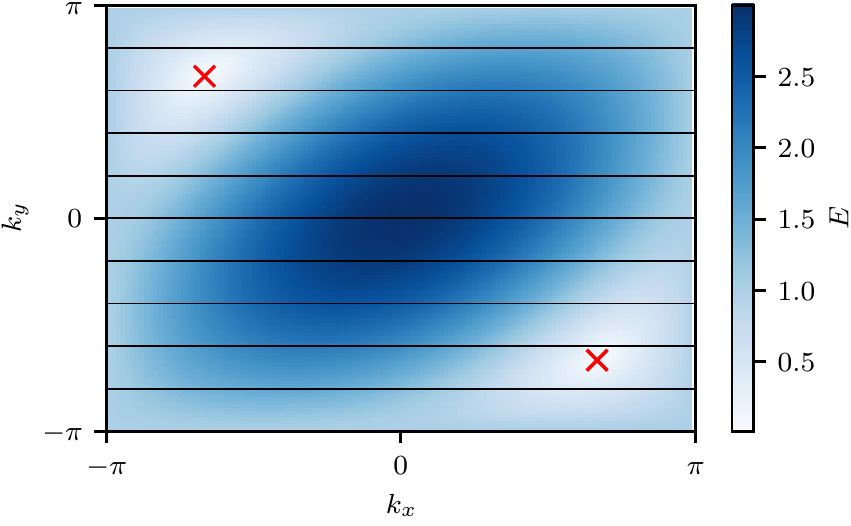}
\caption{The Brillouin zone for a system zig-zag nanotube of $N = 10$ with lattice spacing $a=1$. The horizontal lines are the paths each band $E_n(k_x)$ takes through the BZ. The two red crosses mark the positions of the zero energy Dirac points for an infinite sheet of graphene. We see that, for this particular nanotube, the bands do not pass through the Dirac points, resulting in a gapped system. The $n$th band will pass through a Dirac point, resulting in a gapless system, if $n/N = \pm 1/3$.}
\label{fig:dispersion}
\end{figure}

We impose periodic boundary conditions upon $H$ in the $\boldsymbol{n}_x$ and $\boldsymbol{n}_y$ directions, with $L+1$ and $N$ unit cells in these directions respectively, which gives us a carbon nanotube of length $L$. Therefore, the Hamiltonian can be diagonalised by first taking the discrete Fourier transform
\begin{equation}
a_{\boldsymbol{r}} = \frac{1}{\sqrt{N_c}} \sum_{\boldsymbol{k} \in \text{BZ}} e^{i \boldsymbol{k} \cdot \boldsymbol{r}} a_{\boldsymbol{k}}
\end{equation}
and similarly for $b_{\boldsymbol{r}}$, where $N_c = N(L+1)$ is the number of unit cells in the honeycomb lattice. With this, the Hamiltonian takes the form
\begin{equation}
H = \sum_{\boldsymbol{k} \in \text{BZ}} \psi^\dagger_{\boldsymbol{k}} h(\boldsymbol{k}) \psi_{\boldsymbol{k}}, \ h(\boldsymbol{k}) = \begin{pmatrix} 0 & f(\boldsymbol{k}) \\ f^*(\boldsymbol{k}) & 0 \end{pmatrix}, \label{eq:graphene_ham}
\end{equation}
where $\psi_{\boldsymbol{k}} = (a_{\boldsymbol{k}} , b_{\boldsymbol{k}})^T$ is a two-dimensional spinor, where the sublattice degrees of freedom appear as the ``spin" degrees of freedom of the spinor, and
\begin{equation}
f(\boldsymbol{k}) = -t\left(1 + e^{-i(k_x-k_y)a} + e^{-ik_xa}\right).
\end{equation}
The single-particle dispersion relation of $H$ is given by $E(\boldsymbol{k}) = \pm |f(\boldsymbol{k})|$ which is shown in the 2D colour plot of Fig. \ref{fig:dispersion}. This dispersion is gapless and contains two zero energy Dirac points about which the system acts relativistically, as shown by the crosses in Fig. \ref{fig:dispersion}.
 
To construct a zig-zag nanotube, we let the length $L$ in the $\boldsymbol{n}_x$ direction tend to infinity whilst letting the periodic periodic length in the $\boldsymbol{n}_y$ direction remain finite, with $N$ unit cells in circumference, which creates an infinitely long nanotube. In this case, the Bloch momenta are semi-quantised within the BZ, with $k_x$ unconstrained and 
\begin{equation}
k_y  = \frac{2 n \pi}{Na},
\end{equation}
where $n$ is an integer. The quantisation of $k_y$ means that the system only has access to one-dimensional bands of momentum states within the BZ labelled by the integer $n$ \cite{Charlier}. The $n$th band has the dispersion relation
\begin{equation}
E_{n}(k_x) = \pm t \sqrt{3 + 2g_n(k_x)}, 
\end{equation}
where 
\begin{equation}
g_n(k_x) =  \cos \left( \frac{2 n \pi}{N} \right) +  \cos \left( \frac{2 n \pi}{N} - k_x a \right) +  \cos (k_x a),
\end{equation}
which is obtained by simply substituting the quantised values of $k_y$ into the dispersion $E(\bs k)$ of graphene. For each value of $n$ we have an energy band which is gapped in general, unlike an infinite flat sheet of graphene which is always gapless. This is due to the finite circumference of $N$ unit cells. Each conduction band contains a single minima which will describe the low-energy physics for that particular band. We stress that these minima are not the two zero-energy Dirac points of an infinite sheet of graphene, as these two points are inaccessible to the system in general. Only for special values of $N$ will these points be accessible, yielding a gapless nanotube.

We study the low-energy properties for the $n$th band by Taylor expanding the Hamiltonian about the band minima, following a similar route to that of references \cite{Marganska,Akhmerov}. These minima are located at the same position as the minima of $F_n(k_x) = E_n^2(k_x)$, so we use this as it is easier to work with. The turning points $ K_\text{min}$ obey $F_n'( K_\text{min}) = 0$, which gives us the equation
\begin{equation}
\sin\left(\frac{2n \pi}{N} -  K_\text{min}a \right) = \sin\left( K_\text{min}a \right).
\end{equation} 
Using the result that $\sin(x) = \sin(y) \Rightarrow x = m \pi + (-1)^m y$ for $m \in \mathbb{Z}$, this implies
\begin{equation}
 K_\text{min} = \frac{ \pi}{(1+(-1)^m)a} \left(\frac{2n}{N}-m \right).
\end{equation}
Due to the denominator we have finite solutions only when $m$ is an even number, so we take $m=-2l$ which gives us the turning points 
\begin{equation}
 K_\text{min}  = \frac{\pi}{a} \left( \frac{n}{N} + l \right), \quad l \in \mathbb{Z}.
\end{equation}
Note that as $ K_\text{min} \in [-\pi,\pi]$ and $n/N \in [-1/2,1/2]$, the only possibilities are that $l = -1,0,1$. We now need to identify which of these turning points are minima, where $F''_n( K_\text{min}) > 0$. Substituting in $ K_\text{min}$ from above, we require 
\begin{equation}
(-1)^l \cos\left(\frac{n \pi}{N} \right) < 0.
\end{equation}
As $n/N \in [-1/2, 1/2]$, then $ \cos(\frac{n \pi}{N}) \in [0,1]$, therefore in order to satisfy this constraint whilst ensuring $ K_\text{min} \in [-\pi,\pi]$, we require
\begin{equation}
l = \begin{cases} -1 & \text{if} \ n > 0 \\ 1 & \text{if} \  n \leq 0 \end{cases} \ \Rightarrow \  K_\text{min} = \begin{cases} \left( \frac{n}{N} - 1 \right) \frac{\pi}{a} &   n > 0  \\ \left( \frac{n}{N} + 1 \right) \frac{\pi}{a} &  n \leq  0\end{cases}.
\end{equation}
Each band has a single minima. If $n = N/2$, there does not exist a minima as the band is completely flat so we do not consider this.

The continuum limit Hamiltonian of the $n$th band is defined as $h_n(p) = h(K_\text{min} + p,2n \pi/Na )$ to first order in $p$, where $h(p)$ is given in Eq. (\ref{eq:graphene_ham}). Therefore, substituting in $k_y = 2n \pi/Na$ into $f(\boldsymbol{p})$ gives us the set of $N$ functions
\begin{equation}
f_n(p) \equiv f( p,2n \pi/Na) = -t \left( 1+ e^{-i pa} \left(  e^{i \frac{2 n \pi}{N}}  + 1\right) \right),
\end{equation}
which are enumerated by the band index $n$. We Taylor expand $f_n(p)$ about $K_\text{min}$: 
\begin{equation}
f_n(K_\text{min} + p) = f_n(K_\text{min}) + p f'_n(K_\text{min}) + O(p^2).
\end{equation}
We have
\begin{equation}
\begin{aligned}
f_n(K_\text{min}) & = -t \left( 1 + e^{-iK_\text{min} a} \left(  e^{i \frac{2 n \pi}{N}} +1\right) \right)\\ 
&  = -t \left( 1 + e^{-i \frac{n \pi}{N}} e^{\mp i \pi} \left(  e^{i \frac{2 n \pi}{N}} +1\right) \right) \\
& = -t \left( 1 - e^{-i \frac{n \pi}{N}} \left(  e^{i \frac{2 n \pi}{N}} +1\right) \right) \\
& = -t \left( 1 -  \left( e^{i \frac{n \pi}{N}} + e^{-i \frac{ n \pi}{N}} \right) \right) \\
& = -t \left( 1 - 2 \cos\left( \frac{n \pi}{N} \right) \right) \\
& \equiv \Delta_n.
\end{aligned}
\end{equation}
We also have
\begin{equation}
\begin{aligned}
f'_n(p) & = iat e^{-ipa}\left( e^{i \frac{2 n \pi}{N}} +1\right) \\
\Rightarrow f'_n(K_\text{min}) & =  iat e^{-i K_\text{min}a}\left( e^{i \frac{2 n \pi}{N}} +1\right) \\
& = iat e^{-i \frac{n \pi}{N}} e^{\mp i \pi} \left( e^{i \frac{2 n \pi}{N}} +1\right) \\
& = -iat \left( e^{i \frac{ n \pi}{N}} +e^{-i \frac{ n \pi}{N}} \right) \\
& =-2iat \cos\left( \frac{n \pi}{N}\right) \\
& \equiv -i e_n.
\end{aligned}
\end{equation}
Pulling everything together, we get
\begin{equation}
h_n(p) = \begin{pmatrix} 0 & f_n(K_\text{min} + p) \\ f_n^*(K_\text{min} + p) & 0 \end{pmatrix} = \begin{pmatrix} 0 & \Delta_n -i e_n p \\ \Delta_n + i e_n p & 0 \end{pmatrix} = e_n \sigma^y p + \Delta_n \sigma^x +O(a^2p^2).
\end{equation}
We choose to interpret $e_n$ as a zweibein as it allows us to generalise to curved spacetimes where $t$ is space-dependent. We see that the failure of the continuum limit to describe the flat band $n = N/2$ is encoded in the zweibein as it vanishes here. Taking the limit that $a \rightarrow 0$ whilst keeping $at$ fixed, we can safely ignore the $O(a^2 p^2)$ terms and we now have our continuum limit/low-energy Hamiltonian.

The above Hamiltonian is a $(1+1)$D Dirac Hamiltonian with mass $\Delta_n$ using the representation $\alpha^x =  \sigma^y$ and $\beta = \sigma^x$. The relativistic dispersion relation is given by $E_n(p) = \pm \sqrt{\Delta^2_n + e_n^2 p^2}$. Note the gap closes if $n/N = \pm 1/3$, which is only possible if $N$ is a multiple of three. The corresponding Dirac equation reads
\begin{equation}
 \begin{pmatrix} 0 &   \Delta_n - i e_n p \\   \Delta_n + i e_n p  & 0  \end{pmatrix} \begin{pmatrix} \phi_A \\ \phi_B \end{pmatrix} = E_n \begin{pmatrix} \phi_A \\ \phi_B \end{pmatrix}.
\end{equation}
This yields two equivalent equations, both implying
\begin{equation}
\phi_B = \left( \frac{\Delta_n + i e_n p}{E_n} \right)  \phi_A \equiv s e^{i \theta_{n,p}} \phi_A,
\end{equation}
where $\theta_{n,p} = \arg(\Delta_n +ie_n p)$ and $s = \mathrm{sgn}(E_n)$. The corresponding un-normalised eigenvectors in one-dimensional position space are given by
\begin{equation}
\phi_{n,p}(x) = \begin{pmatrix} 1 \\ s  e^{i \theta_{n,p}} \end{pmatrix}  e^{i px},
\end{equation}
where we now we rename $ax \mapsto x$ which is our continuum coordinate system when $a \rightarrow 0$. We interpret the top and bottom components of our spinors as the wavefunction on sublattices $A$ and $B$ respectively.

With the continuum limit approximation, we now study a nanotube of finite length $L$ in the $\boldsymbol{n}_x$ direction by imposing suitable boundary conditions. Note that for the purposes of numerically encoding this, this requires $L+1$ unit cells in the $\boldsymbol{n}_x$ direction. First, we build positive energy ($s=1$) standing waves by superimposing forward and backward propagating waves as $\psi_{n,p} = \phi_{n,p} + R \phi_{n,-p}$, where $R \in \mathbb{C}$ as
\begin{equation}
\psi_{n,p} =  \begin{pmatrix} 1 \\   e^{i \theta_{n,p}} \end{pmatrix}  e^{i px} +  R \begin{pmatrix} 1 \\   e^{-i \theta_{n,p}} \end{pmatrix}  e^{-i px},
\end{equation}
where $\theta_{n,-p} = -\theta_{n,p}$. The zig-zag boundary conditions are given by $
\psi_A(0) = \psi_B(L) = 0$. These boundary conditions can be seen clearly in Fig. \ref{fig:boundaries} as the unit cells of the top and bottom row, where $x= 0$ and $x=L$, each contain a ``missing" site that is outside of the system (recall that our coordinates $x$ label the unit cell and not the individual sites). The non-relativistic wavefunction must vanish on these sites so the corresponding components of the spinor must vanish. Note that, in our representation of the Dirac alpha and beta matrices, the zero-flux condition of Eq. (\ref{eq:boundary_flux}) reads $\mathrm{Im}(\psi^*_A \psi_B) = 0$ on the boundaries which the zig-zag boundary conditions satisfy. The first boundary condition gives $R = -1$, so our solutions take the form
\begin{equation}
\psi_{n,p}(x) = \mathcal{N}_{n,p} \begin{pmatrix}  \sin(px)  \\ \sin(px + \theta_{n,p}) \end{pmatrix} , 
\end{equation}
where $\mathcal{N}_{n,p}$ is a normalisation constant. The second boundary condition gives $\sin(pL + \theta_{n,p}) = 0$, giving the transcendental equation for the allowed momenta
\begin{equation}
pL + \theta_{n,p} = m \pi , \ m \in \mathbb{N},
\end{equation}
which can be solved numerically by minimising the function $f_{nm}(p) = |pL + \theta_{n,p} - m \pi|$ with respect to $p$ for a fixed $n$,$m$.

\section{Appendix B: Numerical simulation of a zig-zag carbon nanotube}
\begin{figure}
\centering
\includegraphics[width=0.5\columnwidth]{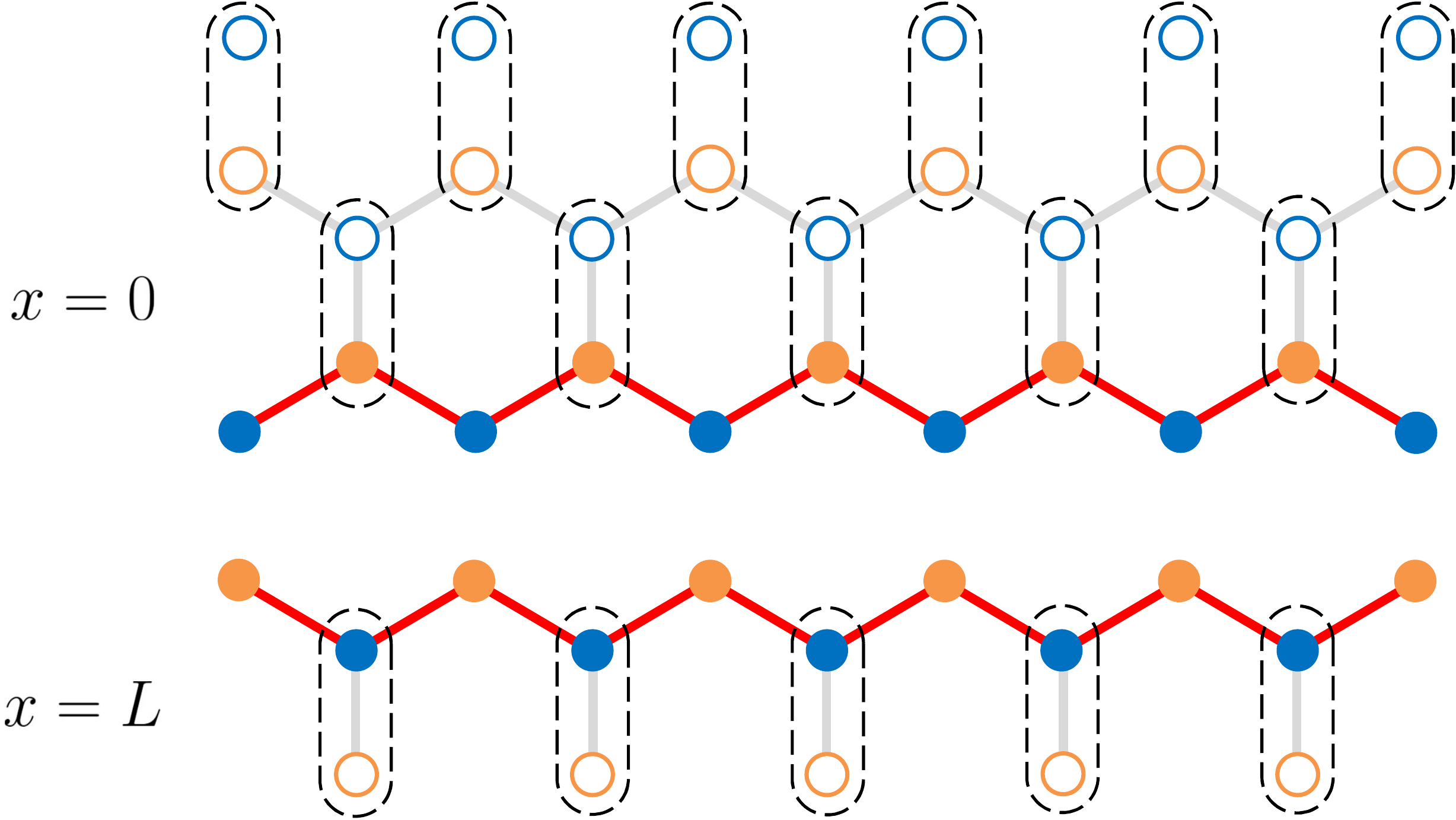}
\caption{Due to our choice of unit cell in the Hamiltonian as seen in Fig. \ref{fig:unit_cell}, the numerical factors $x_{\boldsymbol{r}}$, $y_{\boldsymbol{r}}$ and $z_{\boldsymbol{r}}$ are required to ``switch off" the links that are outside of the system to create the red zig-zag boundaries. At the top boundary where $x=0$, all grey $X$, $Y$ and $Z$ links are outside of the system so the numerical factors are $0$ here. Along the bottom row at $x= L$, only the $Z$ link is switched off.}
\label{fig:boundaries}
\end{figure}

In order to numerically simulate the zig-zag carbon nanotube, we must modify the Hamiltonian Eq. (\ref{eq:ham_tb}) slightly to take into account the open boundaries of the system. We take the Hamiltonian
\begin{equation}
H  =  -t \sum_{{\bf r} \in A} a^\dagger_{\bf r} \left( x_{\bf r} b_{\bs r - {\bs n}_x} + y_{\bs r} b_{\bs r - {\bs n}_x + {\bs n}_y} + z_{\bs r} b_{\bs r}  \right) + \text{h.c.} , \label{eq:ham_tb_tube}
\end{equation}
where $x_{\boldsymbol{r}}, y_{\boldsymbol{r}}, z_{\boldsymbol{r}} \in \{0,1\}$ are numerical factors that take into account the top and bottom boundaries of the system. These terms ``switch off" the external $X$, $Y$ and $Z$ links  of the Hamiltonian respectively, as seen in Fig. \ref{fig:boundaries}, to ensure the nanotube has zig-zag boundaries represented by the red links.

In order to fix which band we are in numerically, we derive the corresponding tight-binding model. We Fourier transform with respect to the $\boldsymbol{n}_y$ direction only, with the definition
\begin{equation}
a_\mathbf{r} =  \frac{1}{\sqrt{N}}  \sum_{p_y} e^{i  p_y y} a_x(p_y).
\end{equation}
Substituting this into the tight-binding Hamiltonian Eq. (\ref{eq:ham_tb_tube}) gives us
\begin{equation}
H = \sum_{p_y} H(p_y),
\end{equation}
where our one-dimensional chains have the Hamiltonian
\begin{equation}
H(p_y) = - t \sum_{i= 0}^L a^\dagger_{i} \left(  z_i b_{i} + ( y_i + x_i e^{ip_y})b_{i-1} \right) + \text{h.c.},
\end{equation}
where we have swtiched to the index $i \in \mathbb{N}$ to label the sites of our one-dimensional chain. 

In order to describe a nanotube, we now set $k_y = 2n \pi/N$ to give us $N$ Hamiltonians $H_n \equiv H(2 n \pi/N)$ which describe each band $n$ of the nanotube. Note that numerically we take the lattice spacing $a=1$. In order to encode this numerically, we need the single-particle Hamiltonian. We can write this Hamiltonian as
\begin{equation}
H_n = \sum_{i,j} a^\dagger_i (h_n)_{ij} b_j + b_j^\dagger (h_n^\dagger)_{ij} a_i,
\end{equation}
where
\begin{equation}
(h_n)_{ij} = - t \left( z_i \delta_{ij} + \left(1 + e^{i \frac{2n \pi}{N}} \right) \delta_{i-1,j} \right).
\end{equation}
Note that $x_i$ and $y_i$ are no longer needed here, as the $\delta_{i-1,j}$ removes the $X$ and $Y$ links on the top of the cylinder as required.

If we define the $2(L+1)$-dimensional spinor $\psi = (a_0,a_1,\ldots a_L ;b_0,b_1,\ldots b_L)^T \equiv (\boldsymbol{a},\boldsymbol{b})^T$, then the many-body Hamiltonian can be written as
\begin{equation}
H_n = \begin{pmatrix} \boldsymbol{a}^\dagger & \boldsymbol{b}^\dagger \end{pmatrix} \begin{pmatrix} 0 & h_n \\ h^\dagger_n & 0 \end{pmatrix} \begin{pmatrix} \boldsymbol{a} \\ \boldsymbol{b} \end{pmatrix} \equiv \psi^\dagger \mathcal{H}_n \psi
\end{equation}
Numerically, we diagonalise the matrix $2(L+1) \times 2(L+1)$-dimensional matrix $\mathcal{H}_n$ which his our single-particle Hamiltonian. In our basis, the first (last) $L+1$ components will be the wavefunctions $\psi_A$ ($\psi_B$) of sublattice $A$ ($B$).

\section{Appendix C: Band-dependence of densities}
\begin{figure}[h]
\includegraphics[width=0.5\columnwidth]{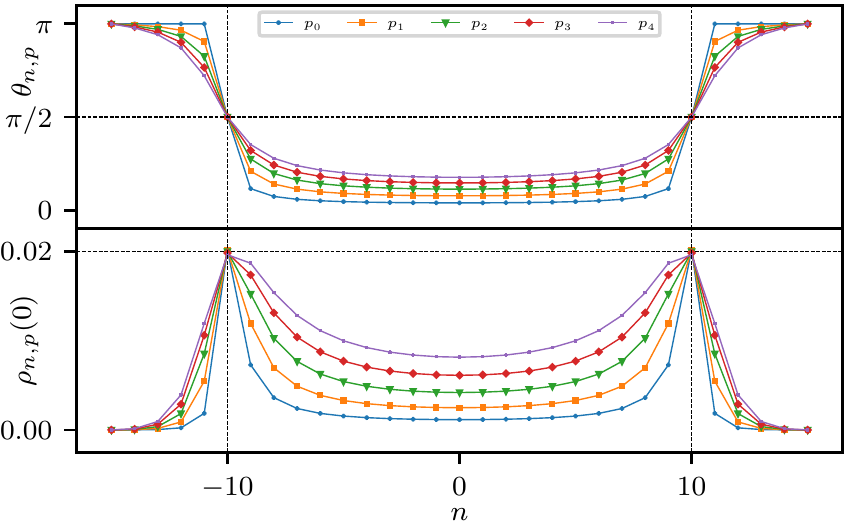}
\caption{ The analytical phase shifts $\theta_{n,p}$ and numerical edge densities $\rho_{n,p}(0) = \rho_{n,p}(L)$ for the ground state and first four excited states of all $N$ bands $n$ of the system of dimension $(N,L) = (30,50)$. We see that the edge density is maximised for bands $n = \pm N/3 = \pm 10$ coinciding with where the phase shift is $\theta_{n,p} = \pi/2$.} 
\label{fig:phases_bands}
\end{figure}

The contrasting behaviour between gapped and gapless systems can also be confirmed numerically by studying each band $n$ of the system, as shown in Fig.~\ref{fig:phases_bands}. In this figure, the edge density of the ground state and first four excited states of a particular band is plotted for all bands $n$ of a system of size $(N,L) = (30,50)$, whereby ground state we mean the lowest energy state of that particular band. We observe that if the nanotube is in an eigenstate of the $n$th band where $n = \pm N/3 =\pm 10$, which are the two gapless bands, then the edge density takes the predicted value of $1/L = 0.02$ for all eigenstates. Away from these special bands a gap opens up, the edge density falls off sharply and a dependence of the density on eigenstate emerges. The edge density behaviour is a consequence of the phase shift, $\theta_{n,p}  = \arg(\Delta_n + ie_np)$, shown in Fig.~\ref{fig:phases_bands}. If $n = \pm N/3 \pm 10$, then $\Delta_n = 0$ exactly, so $\theta_{n,p} = \pi/2$ for all $p$ and consequently all eigenstates behave identically with an edge density of $1/L =0.02$ according to Eq. (10) of the main text. Away from these points the phase rapidly jumps to $0$ or $\pi$, which explains the sharpness of the density peaks. 

\end{document}